# MonALISA : A Distributed Monitoring Service Architecture

H.B. Newman, I.C.Legrand, P. Galvez
*California Institute of Technology, Pasadena, CA 91125, USA*
R. Voicu, C. Cirstoiu
*Polytechnic University Bucharest, Romania*

The MonALISA (Monitoring Agents in A Large Integrated Services Architecture) system provides a distributed monitoring service. MonALISA is based on a scalable Dynamic Distributed Services Architecture which is designed to meet the needs of physics collaborations for monitoring global Grid systems, and is implemented using JINI/JAVA and WSDL/SOAP technologies. The scalability of the system derives from the use of multithreaded Station Servers to host a variety of loosely coupled self-describing dynamic services, the ability of each service to register itself and then to be discovered and used by any other services, or clients that require such information, and the ability of all services and clients subscribing to a set of events (state changes) in the system to be notified automatically. The framework integrates several existing monitoring tools and procedures to collect parameters describing computational nodes, applications and network performance. It has built-in SNMP support and network-performance monitoring algorithms that enable it to monitor end-to-end network performance as well as the performance and state of site facilities in a Grid. MonALISA is currently running around the clock on the US CMS test Grid as well as an increasing number of other sites. It is also being used to monitor the performance and optimize the interconnections among the reflectors in the VRVS system.

## 1. THE MONALISA SERVICES ARCHITECTURE

We are developing a globally scalable ``Dynamic Distributed Services Architecture'' (DDSA) [1], [2] to serve large physics collaborations. This architecture incorporates many features that make it suitable for managing and optimizing workflow through Data Grids composed of hundreds of sites, with thousands of computing and storage elements, and thousands of pending tasks, such as those foreseen by the LHC experiments.

In order to scale and operate robustly in managing global, resource-constrained Grid systems, the DDSA framework uses a set of Station Servers, one per facility or site in a Grid, that host a variety of dynamic, agent-based services. The services are registered with, and can be mutually discovered by a lookup service, and they are notified automatically in case of ``events'' signaling a change of state anywhere in a large distributed system. This allows the ensemble of services to cooperate in real time to gather, disseminate, and process time-dependent state and configuration information about the site facilities, networks, and many jobs running throughout the Grid. The monitored information is reported to higher level services, that in turn analyze the information, and take corrective action to improve the overall efficiency of operation of the Grid (through load balancing, for example) or to mitigate problems as needed. The DDSA framework is inherently distributed, ``loosely coupled'' and self-restarting, making it scalable and robust. Cooperating services and applications are able to access each other seamlessly, to adapt rapidly to a dynamic environment (such as worldwide-distributed analysis by hundreds of physicists in a major HEP experiment). The services are managed by an efficient multithreading engine that schedules and oversees their execution, such that Grid operations are not disrupted if one or more tasks (threads) are unable to continue. The system design also provides reliable ``non-stop'' support for large distributed applications under realistic working conditions, through service replication, and automatic re-activation of services. These mechanisms make the system robust against the failure or inaccessibility of multiple Grid components (when a key network link goes down, for example).

A service in the DDSA framework is a component that interacts autonomously with other services through dynamic proxies or agents that use self-describing protocols. By using dedicated lookup services, a distributed services registry, and the discovery and notification mechanisms, the services are able to access each other seamlessly. The use of dynamic remote event subscription allows a service to register to be notified of a selected set of event types, even if there is no provider to do the notification at registration time. The lookup discovery service will then automatically notify all the subscribed services, when a new service, or a new service attribute, becomes available.

The code mobility paradigm (mobile agents or dynamic proxies) used in the DDSA extends the remote procedure call and the client server approach. Both the code and the appropriate parameters are downloaded dynamically into the system. Several advantages of this paradigm are: optimized asynchronous communication and disconnected operation, remote interaction and adaptability, dynamic parallel execution and autonomous mobility. The combination of the DDSA service features and code mobility makes it possible build an extensible hierarchy of services capable of managing very large Grids, with relatively little program code.

We have built a prototype implementation of the DDSA based on JINI [3] technology. The JINI architecture federates groups of devices and software components into a single, dynamic distributed system; functionality that the future Open Grid Services Architecture (OGSA) [4] will need to include. JINI enables services to find each other on a network and allows these services to participate and cooperate within certain types of operations, while interacting autonomously with clients or other services [5]. This architecture simplifies the construction, operation and administration of complex systems by: (1) allowing registered services to interact in a dynamic and robust (multithreaded) way; (2) allowing the system to adapt when devices or services are added or removed, with no user intervention; (3) providing mechanisms for services to register and describe themselves, so that services can intercommunicate and use other services without prior knowledge of the services' detailed implementation.

We have also included WSDL/SOAP [6], [7] bindings for all the distributed objects, in order to provide access to the monitoring information from other types of clients and to facilitate a possible future migration to the Open Grid Services Architecture

## 2. THE MONITORING SERVICE

An essential part of managing a global Data Grid is a monitoring system that is able to monitor and track the many site facilities, networks, and the many task in progress, in real time. The monitoring information gathered also is essential for developing the required higher level services, and components of the Grid system that provide decision support, and eventually some degree of automated decisions, to help maintain and optimize workflow through the Grid. We therefore developed the agent-based MonALISA (Monitoring Agents in A Large Integrated Services Architecture) [8] system, based on the DDSA framework. MonALISA is an ensemble of autonomous multi-threaded, self-describing agent-based subsystems which are registered as dynamic services and are able to collaborate and cooperate in performing a wide range of monitoring tasks in large scale distributed applications, and to be discovered and used by other services or clients that require such information.

MonALISA is designed to easily integrate existing monitoring tools and procedures and to provide this information in a dynamic, self describing way to any other services or clients. MonALISA services are organized in groups and this attribute is used for registration and discovery.

### 2.1. The Data Collection Engine

The system monitors and tracks site computing farms and network links, routers and switches using SNMP [9], and it dynamically loads modules that make it capable of interfacing existing monitoring applications and tools (e.g. Ganglia [10], MRTG [11], Hawkeye [12]).

The core of the monitoring service is based on a multi-threaded system used to perform the many data collection tasks in parallel, independently. The modules used for collecting different sets of information, or interfacing with other monitoring tools, are dynamically loaded and executed in independent threads. In order to reduce the load on systems running MonALISA, a dynamic pool of threads is created once, and the threads are then reused when a task assigned to a thread is completed. This allows one to run concurrently and independently a large number of monitoring modules, and to dynamically adapt to the load and the response time of the components in the system. If a monitoring task fails or hangs due to I/O errors, the other tasks are not delayed or disrupted, since they are executing in other, independent threads. A dedicated control thread is used to stop properly the threads in case of I/O errors, and to reschedule those tasks that have not been successfully completed. A priority queue is used for the tasks that need to be performed periodically. A schematic view of this mechanism of collecting data is shown in Figure 1.

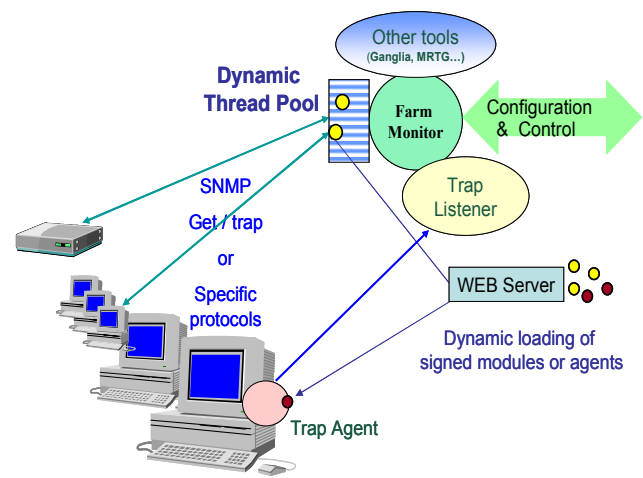

**Figure 1.** A schematic view of the data collection mechanism based on a multi-threaded engine.

This approach makes it relatively easy to monitor a large number of heterogeneous nodes with different response times, and at the same time to handle monitored units which are down or not responding, without affecting the other measurements. As an example, we monitored 500 compute nodes performing a request for ~200 metric values per node every 60 seconds. This provided a sustained rate of ~1600 metric values per second collected, using an average of 20 active threads. The number of threads necessary to monitor a complete site is dynamically adjusted, and very dependent on the response time for each node, which is related to its load as well as to the quality of the network connections.

### 2.2. Data Storage

The collected values are stored in a relational database, locally for each service. The JDBC framework in JAVA offers the flexibility to dynamically load any driver and connect to virtually any relational database. A normalized scheme is used to store the result objects provided by the monitoring modules in indexed tables, which are themselves generated as needed, dynamically. As data are becoming older, we are compressing the values stored in the database by evaluating the mean values on larger time intervals and at the same time keeping the fluctuation range for each parameter.

### 2.3. Registration and Discovery

Each MonALISA service registers with a set of JINI Lookup Discovery Services (LUS) [3] as part of a group, and having a set of attributes. The LUSs are also JINI services and each one may be registered with the other LUSs. If two LUSs have common groups any information related with a change of state detected for a service in the common group by one is replicated to the other one. In this way it is possible to build a distributed and reliable network for registration of services and this technology allows dynamically adding or removing LUSs from the system. Any service should also provide for registration the code base for the proxies that other services or clients need to instantiate for using it. This approach is used to make sure that the right proxies are used for each service while different versions may be used in a distributed organization at the same time. The registration is based on a lease mechanism that is responsible to verify periodically that each service is alive. In case a service fails to renew its lease, it is removed from the LUSs and a notification is sent to all the services or clients that subscribed for such events.

Any monitor client services is using the Lookup Discovery Services to find all the active MonALISA services running as part of one or several group "communities". It is possible to select the services based on a set of matching attributes. The discovery mechanism is used for notification when new services are started or when services are no longer available. The communication between interested services or clients is based on a remote event notification mechanism which also supports subscription.

The client application connects directly with each service it is interested in for receiving monitoring information. To perform this operation, it first downloads the proxies for the service it is interested in from a list of possible URLs specified as an attribute of each service, and than it instantiate the necessary classes to communicate with the service. This procedure allows each service to correctly interact with other services.

### 2.4. Predicates, Filters and Alarm Agents

The clients can get any real-time or historical data by using a predicate mechanism for requesting or subscribing to selected measured values. These predicates are based on regular expressions to match the attribute description of the measured values a client is interested in. They may also be used to impose additional conditions or constrains for selecting the values. In case of requests for historical data, the predicates are used to generate SQL queries into the local database. The subscription requests will create a dedicated thread, to serve each client. This thread will perform the matching test for all the predicates submitted by a client with the measured values in the data flow. The same thread is responsible to send the selected results back to the client as compressed serialized objects. Having an independent thread per client allows sending the information they need, fast, in a reliable way and it is not affected by communication errors which may occur with other clients. In case of communication problems these threads will try to reestablish the connection or to clean-up the subscriptions for a client or a service which is not anymore active.

Monitoring data requests with the predicate mechanism is also possible using the WSDL/SOAP binding from clients or services written in other languages. The class description for predicates and the methods to be used are described in WSDL and any client can create dynamically and instantiate the objects it needs for communication. Currently, the Web Services technology does not provide the functionality to register as a listener and to receive the future measurements a client may want to receive.

Other applications or clients may also use the Agent Filters to receive the information they need. The Agent Filter is a java module which can be dynamically deployed to any MonALISA service, and is design to perform a dedicated data processing task on local data (by subscribing with a predicate to the data flow) and returns back the processed information periodically. The MonALISA service provides the run time environment for these agents which must be digitally signed by a trusted certificate. As an example, such filters are used to compute the aggregate IO traffic in a farm, or to provide the number of nodes which are free. The same thread used for handling the predicate subscription is used for sending the filtered results back to each client.

Dynamically loadable alarm agents, and agents able to take actions when abnormal behavior is detected, are currently being developed to help with managing and improving the working efficiency of the facilities, and the overall Grid system being monitored.

### 2.5. Graphical Clients

We developed a global graphical client which is using the discovery mechanism to find all the active services from a list of user defined groups. This graphical client is developed as a Web Start [13] application and it can be easily started and used from any browser.

A MonALISA service can provide its own GUI to any client as a complex proxy which contains the marshaled components as an attributed to the service [3]. This GUI is used to communicate back with each service from which the user wants detailed information and can plot the requested values. MonALISA provides a flexible access to real-time or historical monitoring values, by using a predicate subscription mechanism or dynamically loadable filter agents. These mechanisms are used by any interested client to query and subscribe to only the information it needs, or to generate specific aggregate values in an appropriate format. When a client subscribes with a predicate to certain values, the GUI will be automatically updated every time a new value matching the subscription is collected.

Graphical user interfaces allow users to visualize global parameters from multiple sites [8], as well as detailed tracking of parameters for any component in the entire system. The graphical clients also use the remote notification mechanism, and are able to dynamically show when new services are started, or when services become unavailable. Dedicated filers are used to provide global views with real time updates for all the running services.

In Figure 2, we present a few examples in how real-time and historical data are presented in MonALISA.

A generic framework for building "pseudo-clients" for the MonALISA services was developed [14]. This has been used for creating dedicated Web service repositories with selected information from specific groups of MonALISA services. The "pseudo-clients" use the same LUSs approach to find all the active MonALISA services from a specified set of groups and subscribes to these services with a list of predicates and filters. These predicates or filters specify the information the pseudo-client wants to collect from all the services. A "pseudo-client" stores all the values received from the running services in a local MySQL database, and is using procedures written as Java threads to compress old data.

A Tomcat [15] based servlet engine is used to provide a flexible way to present global data and to construct on the fly graphical charts for current or customized historical values, on demand. Dedicated servlets are used to generate Wireless Access Protocol (WAP) [16] pages containing the same information for mobile phone users. Multiple Web Repositories can easily be created to globally describe the services running in a distributed environment.

### 2.6. Administration of Services

MonALISA also provides a secure mechanism (SSL with X.509 certificates) for dynamic configuration, using a dedicated GUI, of farms / network elements, and support for other higher level services that aim to manage a distributed set of facilities and/or optimize workflow.

It allows reconfiguring any monitoring services by adding new nodes, network elements or clusters and at the same time to dynamically loaded into the system any new monitoring module as needed. It also allows stopping or suspending any monitoring module. Adding dynamically new monitoring modules is important for debugging and understanding the way certain applications perform.

The Administration interface connects to a service using Remote Method Invocation over SSL. X.509 certificates for trusted administrators are imported in the keystore of each service and they are used to establish a SSL connection based on a client authentification procedure. The administrative GUI can be stated automatically from the global web start client if it used by a trusted

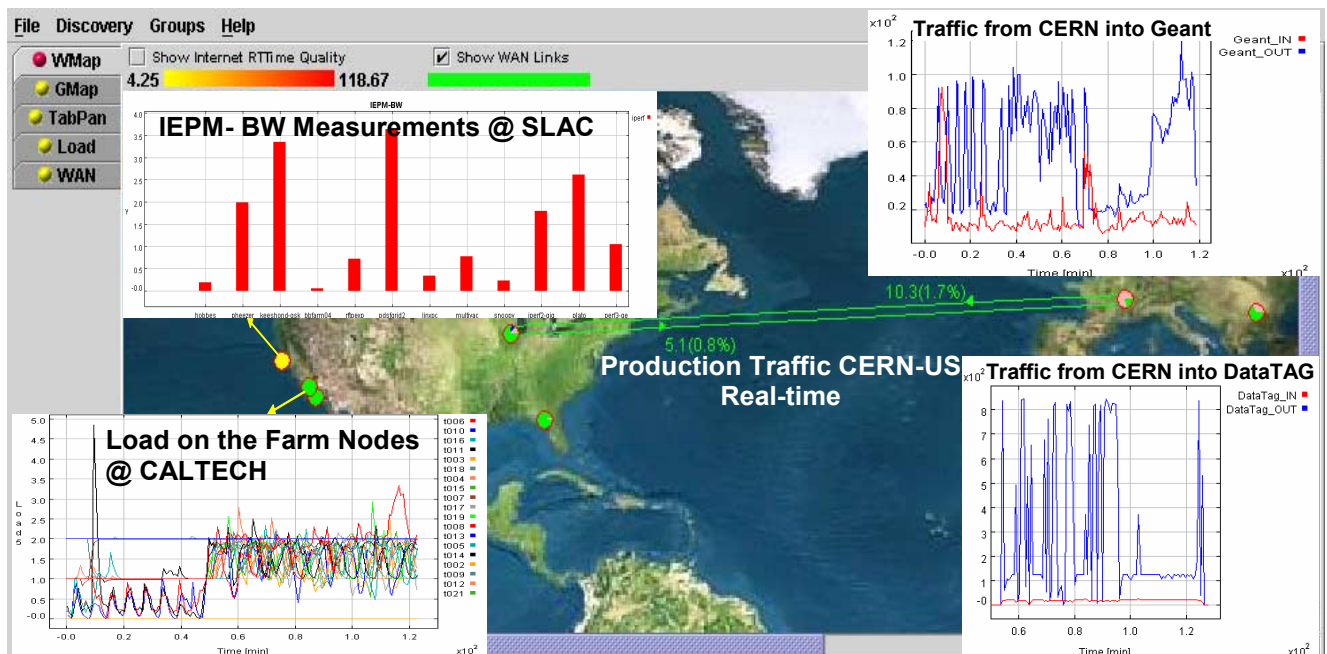

**Figure 2.** The main GUI in MonALISA: it provides global views of the system as well as real time and historical plots for any parameter monitored by the system. Active services are automatically shown on the world map indicating the global load of the farms and real time traffic on selected major international connections. The user can plot any set of parameters measured in the entire system.

**MOET001**

administrator. When the administrator loads is private key into the global GUI client it automatically gets administrative rights on the services that imported his certificate in the trust keystore.

### 2.7. Automatic update for services

MonaALISA is currently deployed on many sites and maintaining and updating such applications may require a significant effort. For this reason we developed a mechanism in MonALISA that allows us to automatically update the monitoring service. A dedicated thread is used to periodically check for updates of the distribution. Alternatively a remote event notification can be used to notify only selected services to perform an update. When such an event is detected, the running service will trigger a restart operation. When a MonALISA service is started, it is using the web start mechanism [13] to describe an application and all its dependencies and constrains into a XML file (jnpl). This will perform an automatic download of all the packages which were updated and will check all the necessary constrains to run the application. All the files downloaded in this way must be digitally signed by a developer for which the certificate is imported in the trust keystore. This can be done when the MonALISA service is used for the first time.

All the running services, as well as the services which may be stated after an update was done will run the last "published" version and this is done in a secure way.
Users may start a MonALISA service with the auto update flag switch off.

### 3. MONITORING DATA PROCESSING FARMS

MonALISA is now deployed and operating round the clock monitoring the US CMS Test Grid and an increasing number of other sites. The MonALISA Web repository is now accumulating historical data for the US CMS Tier1 and Tier2 centers at Fermilab, Caltech, UCSD, and the University of Florida, as well as the production farms at CERN, at the Academia Sinica in Taiwan (ATLAS), and at the Polytechnic University in Bucharest. As an example, the number of nodes loaded on the US-CMS farms during a week is presented in Figure 3.
We also monitor the network traffic on the US-CERN production link, and the distribution of the traffic into the major networks and links with which we peer: EsNet, Abilene, Mren, StarTAP, the US-CERN DataTAG link, the CERN-Geant link, Taiwan-Chicago, and Bucharest-Budapest. In addition to the directed measurements performed on routers, we interfaced MonALISA to provide access to the Internet End to End Performance Measurements (IEPM-BW) [17].
We are currently monitoring the batch queuing systems at CERN (LSF) and at Caltech (PBS). From these modules

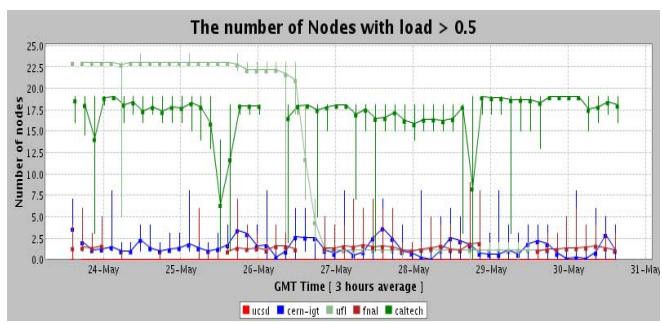

**Figure 3.** A global plot of the US-CMS farms showing the number of nodes with load higher than 0.5 during a period of one week. These plots are created with the web service repository [14].

we can report the number of (selected types) jobs running, pending or those which exit with errors.

### 4. MONITORING THE VRVS SYSTEM

The Virtual Rooms VideoConferencing System (VRVS) [18] is an enhanced web based video conferencing system which is using a set of reflectors distributed world wide for an efficient real-time distribution of the audio and video streams.

For each VRVS reflector, a MonALISA service is running using an embedded Database, for storing the results locally, and runs in a mode that aims to minimize the reflector resources it uses (typically less than 16MB of memory and practically without affecting the system load). Dedicated modules to interact with the VRVS reflectors were developed: to collect information about the topology of the system; to monitor and track the traffic among the reflectors and report communication errors with the peers; and to track the number of clients and active virtual rooms. In addition, overall system information is monitored and reported in real time for each reflector: such as the load, CPU usage, and total traffic in and out.

A dedicated GUI for the VRVS version was developed as a java web-start client. This GUI provides real time information dynamically for all the reflectors which are monitored. If a new reflector is started it will automatically appear in the GUI and its connections to its peers will be shown. Filter agents to compute an exponentially mediated quality factor of each connection are dynamically deployed to every MonALISA service, and they report this information to all active clients who are subscribed to receive this information.

It provides real-time information about the way the VRVS system is used (number of conferences or clients) the topological connectivity of the reflectors and the quality of it and system related information (IO traffic CPU load). Clients can also get historical data for any of these parameters.

The subscription mechanism allows one to monitor in real time any measured parameter in the system as all the updates are dynamically displayed on the open windows. Examples of some of the services and information available, visualizing the number of clients and the active virtual rooms, the traffic in and out of all the reflectors, as well as problems such as lost packets between reflectors are presented in Figure 4.

In addition to dedicated monitoring modules and filters for the VRVS system, we developed agents able to supervise the running of the VRVS reflectors automatically. This will be particularly important when scaling up the VRVS system further.

In case a VRVS reflector stops or does not answer correctly to the monitoring requests, the agent will try to restart it.

If this operation fails twice the Agent will send an email to a list of administrators. These agents are the first generation of modules capable of reacting and taking well defined actions when errors occur in the system. These agents, capable to take action in the system, may be dynamically loaded. For security reasons such agents ping like measurements using UDP packages, which are deployed on all the MonALISA services. These agents perform continuously (every 2s) such measurements with a selected set of possible peers, which can be dynamically reconfigured, for each reflector. We are using small UDP packages to evaluate the Round Trip Time (RTT), its jitter and the percentage of lost packages.

The reflectors and all these possible peer connections we are measuring define a graph (Figure 5). The best routing path for reapplication of the multimedia streams is defined as a Minimum Spanning Tree (MST) [19]. This means that we need to find the tree that contains all the reflectors (vertices in the graph G) for which the total connection "cost" is minimized:

$$MST = \min(\sum_{(v,u) \in G} w((v,u)))$$

The "cost" of the connection between two reflectors (w) is evaluated using the UDP measurements from both sides. This cost function is build with an exponentially mediated RTT and if lost packages are detected or the jitter of the must be digitally signed by developers with trusted certificates, declared for each running service.

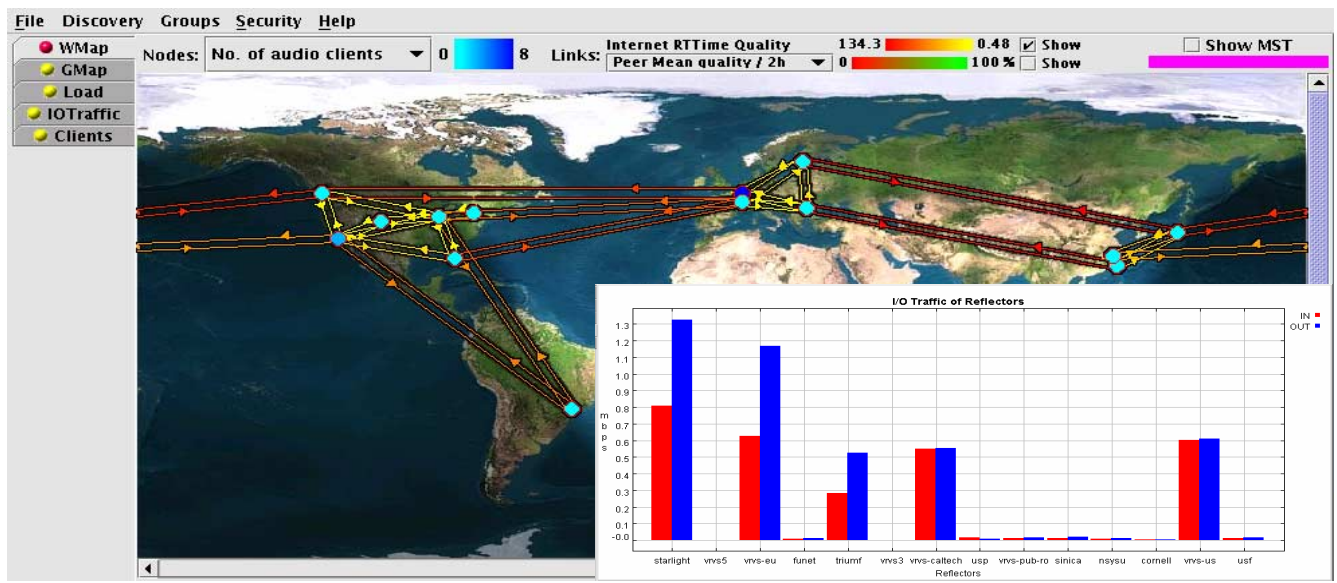

**Figure 4.** Monitoring the VRVS Reflectors.

### 4.1. Optimized Dynamic Routing

We developed agents able to provide an optimized dynamic routing of the videoconferencing data streams. These agents require information about the quality of the alternative connections in the system and they solve, in real-time, a minimum spanning tree problem to optimize the data flow at the global level.

To evaluate the connection quality with possible peer reflectors we developed monitoring agents performing

RTT is high the cost function will increase rapidly.

Based on these values provided by the deployed agents, the MST is calculated nearly in real - time. We implemented the Barůvka's Algorithm [19], as it is well suited for a parallel/distributed implementation. Once a link is part of the MST a momentum factor is attached to that link. This is to avoid triggering reconnections for small fluctuations in the system. Such cases may occur when two possible peers have very similar parameters (or they may be at the same location). In Figure 5 an example of a dynamically MST for connecting the VRVS reflectors is presented.

This is an example of a high level service developed to optimize a real-time world wide distributed application and to help in operating such complex systems. These developments are transforming the VRVS system into a new class of large scale distributed systems with real time constraints.

**Figure 5.** The Minimum Spanning Tree connections and peers quality for a set of VRVS reflectors

The MonALISA framework is a means of carrying out the development of this system, both in terms of its operational characteristics (heuristic, self-discovering, autonomous) and the relatively short development time required for implementing a distributed monitoring and management system of this scale and complexity.

## 5. STATUS AND FUTURE PLANS

Deploying these monitoring services on many sites and interfacing it with other monitoring tools (SNMP, Ganglia, MRTG, IEPM-BW) as well as with batch queuing systems (Condor, LSF, PBS ) has provided very useful experience, and has enabled us to begin building reliable and scalable distributed services.

This experience also has been important in enabling us to start building higher level services, to perform job scheduling and data replication tasks effectively; service that adapt themselves dynamically to respond to changing load patterns in large Grids.

Through the Internet2 End-to-End Performance Initiative [20] MonALISA is also going to be used to monitor and help manage the Internet2 Abilene backbone. We are working to enhance the end-to-end measurements provided by MonALISA to meet the needs of Internet2, as well as the proposed UltraLight next-generation optical network [21].

## 6. SUMMARY

These developments have a broader range of applications, to the global distributed Grid-based systems

MOET001

required for major HENP experiments, and other data-intensive project. This real time system also includes much of the functionality required of the OGSA standardized services planned by the Global Grid Forum in the future.

Effective and robust integrated applications require higher level service components able to adapt to a wide range of requests, and changes in the state of the system (such as changes in the available resources, for example).

These services should be capable of ``learning" from previous experience, and apply ``self-organizing neural network" [22] or other heuristic algorithms to the information gathered, to optimize dynamically the system, by minimizing a set of ``cost functions".

## Acknowledgments

The authors wish to thank to S. Ravot, S. Singh and V. Litvin form Caltech, Richard J. Cavanaugh from University of Florida, Lothar Bauerdick, Ian Fisk, Greg Graham and Yujun Wu from Fermilab, N. Tapus from the Polytechnic University Bucharest, Les Cottrel and Warren Matthews from SLAC for their help and support in deploying and developing MonALISA as a real distributed service.